\documentclass[twocolumn,showpacs,preprintnumbers,amsmath,amssymb]{revtex4}

\usepackage{graphicx}
\usepackage{dcolumn}
\usepackage{bm}

\begin{document}


\title{Single-Shot Readout of Macroscopic Quantum Superposition State in a Superconducting Flux Qubit}
\author{Hirotaka Tanaka$^{1,3}$, Shiro Saito$^{1,3}$, Hayato Nakano$^{1,3}$, Kouichi Semba$^{1,3}$,\\
Masahito Ueda$^{1,2,3}$ and Hideaki Takayanagi$^{1,3}$}

\affiliation{$^{1}$NTT Basic Research Laboratories, NTT Corporation, Atsugi, Kanagawa, Japan\\
$^{2}$Department of Physics, Tokyo Institute of Technology, Tokyo, Japan\\
$^{3}$CREST, Japan Science and Technology Agency, Japan}
\date{July 7, 2004}
             
\begin{abstract}
Single-shot readout experiments were performed on the two lowest-energy states of a superconducting qubit with three Josephson junctions embedded in a superconducting loop. We measured the qubit state via switching current ($I_{\rm sw}$) of a current-biased dc-SQUID, a quantum detector surrounding the qubit loop. The qubit signals were measured in a small $I_{\rm sw}$ regime of the SQUID, typically less than 100 nA, where the $I_{\rm sw}$ distribution is particularly narrow. The obtained single-shot data indicate that the qubit state is readout, through the flux generated by the qubit persistent-current, as energy eigenstates rather than current eigenstates. 
\end{abstract}

\pacs{03.67.Lx, 85.25.Dq, 85.25.Cp, 03.65.Yz}
\maketitle

Among many candidates toward realization of quantum computation, superconducting qubits based on Josephson junctions take on a position of increasing importance \cite{NC-book00}. Although their coherence time still need to be substantially improved, solid state qubits such as semiconductor qubits or superconductor qubits have the distinct merit of scalability due to mature state of nanometer-scale fabrication technology. Recently, several groups have achieved a fairly long coherence time, and gate operation for single qubits has become available by the application of custom made resonant microwave pulses, which control Rabi oscillations \cite{Yasu-Nature99,VACJPUED-Sci02,YCCW-Sci02,MNAU-PRL02,CNHM-Sci03,PYANAT-Nature03}. Rotational gate operation together with two-qubit controlled NOT operation has also been successfully demonstrated \cite{YPANT-Nature03}. As regards the qubit readout, averaging processes are commonly used and a way to overcome imperfect readout visibility has been a topic of debate. 

In this paper, we report the observation of the single-shot readout of a superconducting flux qubit comprising three Josephson junctions in a loop \cite{MOLTWL-Sci99}. Although, single qubit operation and entanglement between two qubits have been demonstrated in superconducting charge qubits, fabricated Josephson circuits exhibit static and dynamic charge noise due to background charged impurities. In contrast, in flux qubits there is relatively little magnetic background noise. Superconducting flux qubits therefore have the potential advantages of a longer coherence time and greater stability. 

\begin{figure}
\includegraphics[width=8cm]{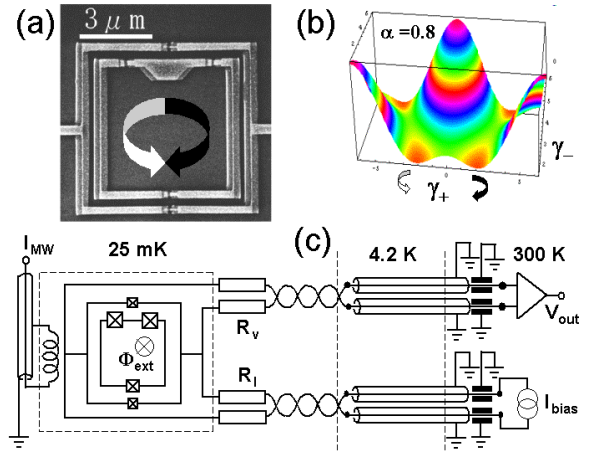}
\caption{\label{fig:sample} 
\footnotesize
(a)	 Scanning electron microscope image of a superconducting qubit (inside loop) and a dc-SQUID (outside loop) as a detector.  Arrows indicate the directions of the supercurrent in the qubit.
(b)	Bird's-eye view of the Josephson energy of the qubit in phase space.
(c)	Schematic drawings of the measurement circuit. The squares with crosses represent Josephson junctions.}
\end{figure}

Figure 1(a) shows a scanning electron microscope image of a sample; a superconducting qubit (inside loop with three Josephson junctions of critical current $I_{\rm c}^{\rm qubit}\simeq 0.6 \mu$A for larger junctions) and an under-damped dc-SQUID (outside loop with two small Josephson junctions of critical current $I_{\rm c}^{\rm SQ}\simeq 0.15 \mu$A for each junction) as a detector. The qubit contains three Josephson junctions, two of which have the same Josephson coupling energy $E_{\rm J}=\hbar I_{\rm c}/2e$. The third has ${\alpha}E_{\rm J}$, with $0.5<\alpha<{1}$. The $\alpha$ value can be controlled by the ratio of the area of the smallest Josephson junction to the other two larger junctions in the qubit. We used a sample with $\alpha=0.8$ with the areas of the larger and smaller junctions given by 0.1$\times$0.4 $\mu$m$^{2}$ and 0.1$\times$0.32 $\mu$m$^{2}$, respectively. The loop area ratio of the qubit to the SQUID was about 3:5. 
Arrows represent macroscopic supercurrent flows in the qubit. The dc-SQUID with two Josephson junctions was under-damped (no shunt resistor). The qubit and the dc-SQUID were coupled magnetically via mutual inductance $M\simeq7$ pH. The aluminum Josephson junctions were fabricated by suspended bridge and shadow evaporation techniques \cite{D-APL77}. The sample was cooled to 25 mK with a dilution refrigerator and it underwent a superconducting transition at $\sim$1.2 K. In order to reduce external magnetic field fluctuations, both the sample holder and the operating magnet were mounted inside a three-fold $\mu$-metal can in the refrigerator. Figure 1(b) shows the qubit potential energy $U(\gamma_{+},\gamma_{-})$ in the phase space, where 
\begin{equation}
\frac{U}{E_{\rm J}}=2+\alpha-\cos{\gamma_{1}}-\cos{\gamma_{2}}-\alpha\cos({2\pi f-\gamma_{1}-\gamma_{2}}), 
\end{equation}
and $\gamma_{1}+\gamma_{2}+\gamma_{3}=2\pi f$, with $\gamma_{i}$ being the phase difference across the three junctions \cite{Orando_PRB}. Junctions 1 and 2 have identical $E_{\rm J}$, but junction 3 with $\gamma_{3}$ has $\alpha E_{\rm J}$. The filling $f$ is the external flux $\Phi_{\rm ext}$ penetrating the qubit loop normalized by the superconducting flux quantum $\Phi_{0}=\frac{h}{2e}$. In a low-energy approximation, the qubit potential can be described by $\gamma_{+}$ alone, where $\gamma_{\pm}\equiv(\gamma_{1}\pm\gamma_{2})/2$. 

\begin{figure}
\includegraphics[width=9cm]{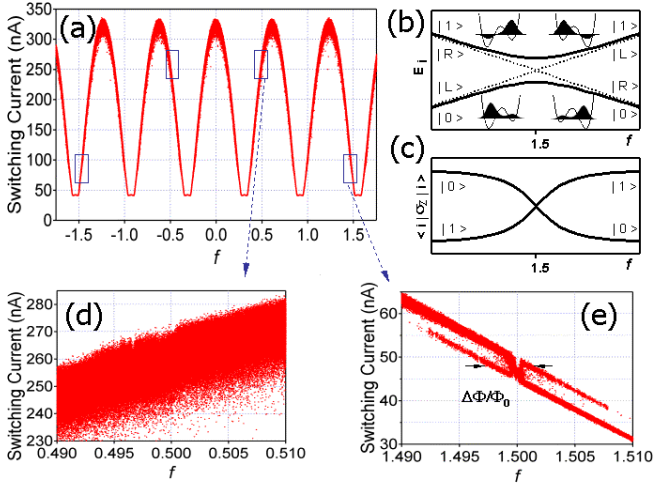}
\caption{\label{fig:sample}
\footnotesize
(a)	Switching current of dc-SQUID detector. $f=\Phi_{\rm ext}/\Phi_{0}$, where $\Phi_{\rm ext}$ is the external magnetic flux penetrating the qubit loop.(b) Schematic illustration of the qubit energy dispersion relation. (c) Expected readout for the qubit energy eigenstates $|0\rangle$ and $|1\rangle$. (d)	Enlarged view near $f=0.5$. (e)	Enlarged view near $f=1.5$. The two states of the qubit are read out distinctly.
}
\end{figure} 

By carefully designing the junction parameters  \cite{MOLTWL-Sci99, Wal}, 
the inner loop can be made to behave as an effective two-state system  \cite{Saito,Nakano}. 
In fact, the read-out result of the qubit changes greatly with qubit design, ranging from the purely classical to the quantum regime 
 \cite{Takayanagi}. The qubit is described by the Hamiltonian 
$H_{\mathrm{q}}=\frac{1}{2}(\varepsilon_{f} \hat{\sigma}_{z}+\Delta \hat{\sigma}_{x})$, where $ \hat{\sigma}_{x,z}$ are the Pauli matrices. Two eigenstates of $\hat{\sigma}_{z}$ are localized states with the supercurrent circulating in opposite directions, i.e., the clockwise state $|R\rangle$ and the counter-clockwise state $|L\rangle$. The dc-SQUID picks up a signal that is proportional to $\hat{\sigma}_{z}$.  
\begin{figure}
\includegraphics[width=9cm]{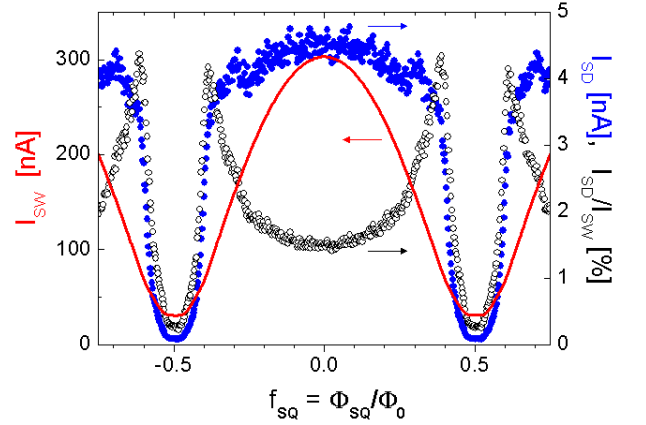}
\caption{\label{fig:sample}
\footnotesize 
DC-SQUID switching current ($I_{\rm sw}$) distribution as a function of $f_{\rm SQ}=\Phi_{\rm SQ}/\Phi_{0}$, where $\Phi_{\rm SQ}$ is the external magnetic flux penetrating the SQUID loop. The standard deviation of the measured switching-current $I_{SD}$ ($\bullet$) and its ratio to $I_{\rm sw}$; $I_{SD}/I_{\rm sw}$ ($\circ$) are plotted. Note that $I_{SD}$ suddenly drops in regions where $I_{\rm sw}$ is smaller than $\sim$100 nA.}
\end{figure}
The barrier for quantum tunneling between these states depends strongly on the Josephson energy $E_{\rm J}$, charging energy $E_{\rm c}$ and $\alpha$ values. The tunneling matrix element is given as 
\begin{equation}
t_{LR}=\frac{\Delta}{2}\approx1.3\sqrt{E_{\rm c}E_{\rm J}}\exp(-0.64\sqrt{\frac{E_{\rm J}}{E_{\rm c}}})
\end{equation}
for $\alpha=0.8$. In the present sample, $E_{\rm J}=1.27$ meV and $E_{\rm J}/E_{\rm c}=50$. 
As schematically shown in Fig. 1(c), the qubit is biased with a static magnetic flux $\Phi_{\rm ext}$ using an external coil. A microwave antenna was placed just above the sample chip to induce oscillating magnetic fields in the qubit loop. The switching voltage of the SQUID was measured by the four-probe method. The detector SQUID was current biased through resistors $R_{I}$ to avoid the parasitic capacitance of the leads. The measurement lines were composed of a pair of resistors ($R_{I}=R_{V}\approx 200 \Omega$), constantan twisted wires up to the mixing chamber position, flexible coaxial cables from the mixing chamber up to the top flange and 10 nF through capacitors at room temperature. The SQUID bias current ($I_{\rm b}$) in a saw-tooth wave was applied with a repetition rate of few hundred Hz. The amplitude of $I_{\rm b}$, typically up to a few hundred nA, depends on the qubit operating point. The SQUID switching voltage was amplified by a differential amplifier and ($I_{\rm sw}, V_{\rm sw}$) switching was recorded when a voltage greater than the threshold ($\sim 30\mu$V) was first detected during every $I_{\rm b}$ ramp.

Figure 2(a) shows the modulation of the measured dc-SQUID switching current against the external magnetic flux as a function of $f=\Phi_{\rm ext}/\Phi_{0}$. Each dot in the graph corresponds to a single-shot readout without averaging. The classically stable current eigenstates $|L\rangle$ and $|R\rangle$ are no longer stable in the presence of quantum tunneling $t_{LR}$. As shown in Fig. 2(b), the qubit ground state $|0\rangle$ and the first excited state $|1\rangle$ change the direction of the persistent current as a function of $f$. Here, 
\begin{eqnarray}
|0\rangle &=& \sin(\frac{\theta_{f}}{2})|L\rangle+\cos(\frac{\theta_{f}}{2})|R\rangle\\
|1\rangle &=&-\cos(\frac{\theta_{f}}{2})|L\rangle+\sin(\frac{\theta_{f}}{2})|R\rangle,
\end{eqnarray}
 where $\theta_{f}=\arctan(\Delta/\varepsilon_{f})$.
Thus, the energy eigenstates $|0\rangle$ and $|1\rangle$ are superpositions of macroscopically distinct states $|L\rangle$ and $|R\rangle$ \cite{Leggett,MS+S2002}. The expected qubit response in $\sigma_{z}$ is shown schematically in Fig. 2(c). Therefore, this step-like feature, which is called a ``qubit step", is expected to occur near $f\approx\frac{1}{2}+n$, where $n$ is an integer (indicated by the boxes in Fig. 2(a)).
Figure 2(d) shows an enlarged view near $f=0.5$, where the switching current distribution of the SQUID is very broad. It is almost impossible to distinguish between the two states of the qubit {$|0\rangle$} and {$|1\rangle$} by a single-shot measurement. In Fig. 2(e) near $f=1.5$, the switching current distribution of the SQUID is narrow enough for us to observe the $\chi$-shaped qubit steps clearly. The observed amount of magnetic-flux shift $\Delta \Phi$ shown in Fig. 2(e)  is consistent with the expected value $2\alpha M I_{\rm c}^{\rm qubit}=3.4\times 10^{-3}\Phi_{0}$. Here, we are able to observe the two energy eigenstates of the qubit with a single-shot measurement. 

\begin{figure}
\includegraphics[width=6cm]{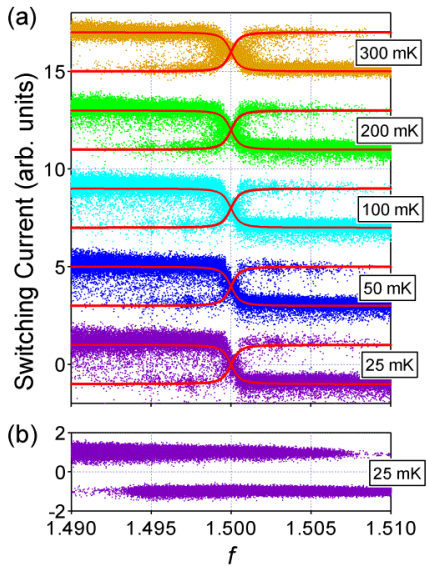}
\caption{\label{fig:sample}
\footnotesize 
(a) Normalized qubit steps as a function of $f=\Phi_{\rm ext}/\Phi_{0}$, where $\Phi_{\rm ext}$ is the external magnetic flux penetrating the qubit loop.
Temperature-averaged expected qubit responses are also shown as solid curves, from which the sinusoidal background modulation of the SQUID has been subtracted. The step height is normalized at 2. Each curve is shifted vertically for clarity.
(b) The classical behavior of a two-level system observed in a different sample.}
\end{figure}

Figure 3 shows the dc-SQUID switching current distribution together with the average switching current as a function of $f_{\rm SQ}=\Phi_{\rm SQ}/\Phi_{0}$, where $\Phi_{\rm SQ}$ is the external magnetic flux penetrating the SQUID loop. The sample used here is nominally identical to that of Fig. 2 and shows a similar $\chi$-shaped crossing. We adopted the standard deviation as a measure of the switching current distribution. We obtained a typical external magnetic flux dependence $I_{\rm sw}\propto|\cos{\frac{\pi\Phi_{\rm SQ}}{\Phi_{0}}}|$. For larger $I_{\rm sw}$ values, the standard deviation of the measured switching-current $I_{SD}$ ($\bullet$) remains large ($\sim 4$ nA) due to stochastic switching processes, which can be described by the macroscopic quantum tunneling (MQT) of the ``phase-ball" in the tilted Josephson washboard potential. The flux detection performance of the SQUID is determined by the distribution and sensitivity of the detector dc-SQUID. The flux sensitivity of the SQUID is proportional to the slope of the $I_{\rm sw}$ modulation curve, i.e., $|\frac{dI_{\rm sw}}{d\Phi_{\rm SQ}}|$ and becomes higher near $f_{\rm SQ}\approx\frac{1}{2}+m$, where $m$ is an integer. In addition, the switching current distribution becomes particularly narrow in the small critical current region, where $I_{\rm sw}$ is typically smaller than 100 nA. This region corresponds exactly to the crossover region $E_{\rm J}/E_{\rm c}\sim 1$ of the detector dc-SQUID, since the area of each Josephson junction in the dc-SQUID is 0.1$\mu$m$\times$0.1$\mu$m, which is roughly four times smaller than that of the qubit junction, resulting in $E_{\rm J}/E_{\rm c}\approx 50/4^{2}\simeq 3$ near $f_{\rm SQ}\simeq 0$, where $I_{\rm sw}\simeq 300$ nA; therefore $E_{\rm J}/E_{\rm c} < 1$ when $I_{\rm sw}$ is smaller than $\sim$ 100 nA. It is notable that a precipitous change in the signal-to-noise ratio occurs at $E_{\rm J}/E_{\rm c}\sim 1$. 
Another possibility is dissipation induced narrowing. The ground state wavefunction of the SQUID localizes more strongly by the incoming noise, resulting in a narrower $I_{\rm sw}$ distribution\cite{MDC-PRB1987}. For small-capacitance low-critical current Josephson junctions, it is known that the width of switching histogram decreses as a consequence of activation process due to effective temperature over the dissipation barrier \cite{Vion-PRL1996}.
Consequently, we have chosen to use this extremely sensitive region of the SQUID detector. By choosing appropriate integers m and n when designing the qubit and SQUID, we succeeded in tuning the qubit operating point, $f=1.5$, to the high sensitivity region of the SQUID. 

\begin{figure}
\includegraphics[width=6cm]{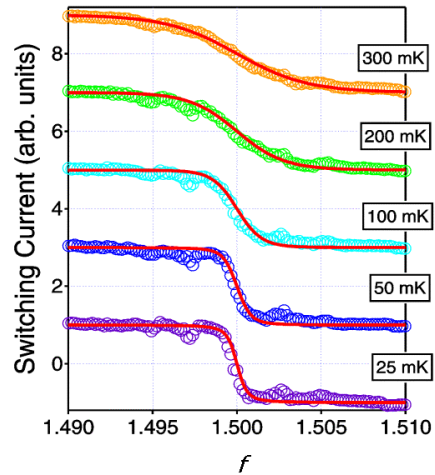}
\caption{\label{fig:sample}
\footnotesize 
Normalized dc-SQUID switching current readout (open circles) after taking an average at each flux bias.  The solid curves represent thermal-averaged theoretical values. 
}
\end{figure}

By subtracting the background modulation of the dc-SQUID, we can extract the signals from the qubit as shown in Fig. 4(a). The normalized switching current of the dc-SQUID against the external magnetic flux is shown as a function of $f=\Phi_{\rm ext}/\Phi_{0}$ at different temperatures. Each dot in the figure corresponds to a single measurement and no averaging is performed. The solid curves are fitted to the ground and excited states with 
\begin{eqnarray}
\langle 0|{\hat \sigma_{z}}|0\rangle &=&-\frac{\varepsilon_{f}}{\sqrt{{\varepsilon_{f}}^2+\Delta^2}}\\
\langle 1|{\hat \sigma_{z}}|1\rangle &=& \frac{\varepsilon_{f}}{\sqrt{{\varepsilon_{f}}^2+\Delta^2}}, 
\end{eqnarray}
where $\varepsilon_{f}=2I_{\rm p}\Phi_{0}(f-1.5)$ with $I_{p}\simeq I_{\rm c}^{\rm qubit}\sqrt{1-(\frac{1}{2\alpha})^2}$. The thicker stripe corresponds to the ground state $|0\rangle$ and the other stripe corresponds to the excited state $|1\rangle$. Here, we should point out that we can expect the dc-SQUID to detect $|L\rangle$ or $|R\rangle$ , i.e., the current eigenstate, by a single-shot readout, in the same way as the Stern-Gerlach apparatus. However, the state observed by this single-shot measurement is the energy eigenstate not the current eigenstate of the qubit. Recently, we have theoretically investigated the switching current behavior as a function of the external magnetic flux $\Phi_{\rm qubit}/\Phi_{0}$ with the conditions of our measurement method\cite{Nakano}. We found that the qubit energy eigenstate of superposition between $|{\rm L}\rangle$ and $|{\rm R}\rangle$ is maintained until the readout, i.e., SQUID switching event, if the relaxation time is long enough and qubit-SQUID interaction is weak as is in our experiments. Whereas, in the superconducting charge qubit, single-shot detection on a non-energy eigenstate basis has recently been reported \cite{Astafiev-PRB2004}. We recorded two thousand switching events under each fixed external magnetic field. The measurement results emerge randomly on either stripe and there is only a small fraction of points in between. Several events were detected in the excited states at the lowest temperatures, which may be due to unexpected excitation related to certain cavity modes. 

For comparison, Fig.4(b) shows an example of a classical two-level system observed in a different sample. The sample parameters are $E_{\rm J}/E_{\rm c}=185$, $E_{\rm J}=2.4$ meV, $\alpha=0.8$, with a calculated qubit energy gap $\Delta=0.2$ $\mu$eV. In this sample, the readouts for the two states are completely separated as $|L\rangle$ or $|R\rangle$ by a single shot measurement. Moreover, the readout does not contain $\chi$ shaped crossing qubit steps. This means that the tunneling term $\Delta$ of the qubit is so small that decoherence easily destroys the coherent tunneling between the wells. As the MQT is suppressed by the high tunnel barrier, only thermal excitation or incoherent tunneling is possible and, consequently, classical bi-stable branches are observed.

\begin{figure}
\includegraphics[width=7.5cm]{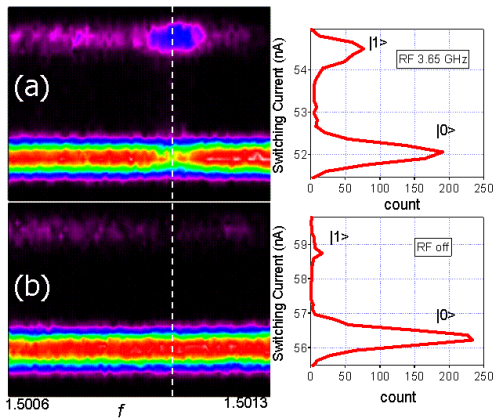}
\caption{\label{fig:sample} 
\footnotesize
A precise readout distribution measurement of the region indicated in Fig. 4(a) (25 mK) (a) with and (b) without microwave irradiation. This sample is from the same set of samples showing the $\chi$-shaped crossing used in Fig. 4(a). The occupancy of the excited state, {$|1\rangle$}, increases with a microwave at a resonant point. The corresponding histograms at the resonance point, $f=1.50102$, are shown in the right panels.}
\end{figure}

If we take the average at each flux bias $f$, the steep step observed at lower temperatures becomes rounded at higher temperatures as expected. Figure 5 shows, the normalized readout after averaging (open circles). This averaging corresponds to time ensemble averaging. Thermally averaged expectation values shown as solid curves in Fig. 5 are calculated from the canonical distribution :  
\begin{equation}
\langle \delta I_{sw} \rangle=(\sum_{i=0}^{1} e^{-\beta E_{i}}\langle i|{\hat \sigma_{z}}|i\rangle) / (\sum_{i=0}^{1} e^{-\beta E_{i}}),
\end{equation}
 where $E_{0(1)}=-(+)\frac{1}{2}\sqrt{{\varepsilon_{f}}^2+\Delta^2}$, and $\beta^{-1}=k_{\rm B}T$ is the effective temperature, $k_{\rm B}$ is the Boltzmann constant. We obtained reasonable agreement between the experimental and calculated results. This is an example of the direct observation of the thermal distribution of a coherent macroscopic quantum object. This shows that even a macroscopic quantum object obeys canonical distribution as if it were a microscopic object. 
We observed two small dips and two small peaks in the averaged qubit step especially at lower temperatures. These dips and peaks appeared in a symmetrical position and could be due to unexpected resonant modes in the environment formed by the sample and sample holder. 

Figure 6 shows a density plot and a histogram of the switching current of the sample showing the $\chi$ cross with and without a microwave. The left panel shows an enlarged view of the region indicated by the box in Fig. 4(a) at 25 mK, and the right panel shows a histogram at the resonance point, $f=1.50102$. The readout is clearly separated into {$|0\rangle$} and {$|1\rangle$} states even with a continuous microwave. Without a microwave the readout is largely in the {$|0\rangle$} state. The residual probability in the {$|1\rangle$} state is due to incoherent thermal excitations and is very small. In contrast, we found that the probability appeared in the {$|1\rangle$} state with microwave irradiation. Ideally, the probabilities of detecting these two states should be the same. 
A microwave transition enables us to estimate the coherence time in the system. 
$T^{\ast}_{2}\simeq 5$ ns has been obtained from the line width of the transition under a resonant microwave.  
The sample showing no $\chi$ cross (Fig. 4(b)) did not respond to the microwave at all, while a coherent transition by the resonant microwave was observed for the sample showing $\chi$ cross (Fig. 4(a)). The observation of the transition induced by the microwave indicates the existence of macroscopic quantum coherence in the system. This is a consequence of a tunneling term $\frac{\Delta}{2}$ in the Hamiltonian. Therefore, in addition to the $\chi$ crossing qubit step, the observation of a microwave resonant transition provides additional evidence for the coherent superposition of macroscopic supercurrent states {$|L\rangle$} and {$|R\rangle$}. 

In conclusion, we reported on the single-shot measurement of a macroscopic quantum state in a superconductor qubit. We achieved a high-resolution readout by carefully selecting the flux bias point and taking advantage of the highly sensitive switching current region of the detector dc-SQUID. We believe that single quantum event detection in fabricated quantum circuits will contribute to a further understanding of the quantum/classical boundary including quantum measurement processes, powered by the recent rapid development of nano-technology. 

We thank J. E. Mooij for motivating this study. We also thank C. J. P. M. Harmans, C. van der Wal, Y. Nakamura, M. Devoret and D. Vion for fruitful discussions, Y. Sekine, K. Shan Kivong, M. C. Goorden, S. Kohlar and E. Laffosse for sample fabrication. The measurement system was partly developed by P. Delsing and T. Claeson. 
 

\end{document}